\documentclass[prl,showpacs,superscriptaddress,twocolumn]{revtex4-1}
\usepackage{amsmath}
\usepackage{amsfonts}
\usepackage{amssymb}
\usepackage{graphicx}
\usepackage{natbib}

\usepackage{color}
\usepackage{array,tabularx}

\def\<{\langle}
\def\>{\rangle}

\begin{document}

\title{Force Percolation Transition of Jammed Granular Systems}

\author{Sudhir N. Pathak}
\affiliation{
Division of Physics and Applied Physics, School of Physical and Mathematical Sciences, 
Nanyang Technological University, Singapore
}
\author{Valentina Esposito}
\affiliation{
Dipartimento di Matematica e Fisica - Universit\`a degli studi della Campania ``Luigi Vanvitelli'', Viale Lincoln 5, 81100 Caserta, Italy
}
\author{Antonio Coniglio}
\affiliation{
CNR--SPIN, Dipartimento di Scienze Fisiche,
Universit\`a di Napoli Federico II, I-80126, Napoli, Italy
}
\author{Massimo Pica Ciamarra}\email[]{massimo@ntu.edu.sg}
\affiliation{
Division of Physics and Applied Physics, School of Physical and Mathematical Sciences, 
Nanyang Technological University, Singapore
}
\affiliation{
CNR--SPIN, Dipartimento di Scienze Fisiche,
Universit\`a di Napoli Federico II, I-80126, Napoli, Italy
}

\date{\today}

\begin{abstract}
The mechanical and transport properties of jammed materials originate from an underlying 
percolating network of contact forces between the grains.
Using extensive simulations we investigate the force-percolation transition of this network,
where two particles are considered as linked if their interparticle force overcomes a threshold.
We show that this transition belongs to the random percolation universality class, thus
ruling out the existence of long-range correlations between the forces.
Through a combined size and pressure scaling for the percolative quantities,
we show that the continuous force percolation transition evolves
into the discontinuous jamming transition in the zero pressure limit, 
as the size of the critical region scales with the pressure.
\end{abstract}
\pacs{64.60.Av, 83.80.Fg, 61.43.Er, 62.20.F, 61.20.Ja}
\maketitle

\begin{figure}[t!!!!]
\setlength{\unitlength}{1cm}
\begin{picture}(8,5)
\thicklines
\put(0.5,0.4){\vector(0,1){4.5}}
\put(0.5,0.4){\vector(1,0){7.5}}
\put(1.5,0.4){\line(3,2){6}}
\put(1.5, 0.4){\circle*{0.2}}
\thinlines
\put(7.3,0.0){$\Phi$}
\put(0,4.3){$f_t$}
\put(1.4,0.0){$\Phi_J$}
\put(6.1,4.2){$f_c(\Phi)$}
{
\setlength{\fboxsep}{0pt}%
\setlength{\fboxrule}{1pt}
\put(2,2.4){\fbox{\includegraphics*[scale=0.13]{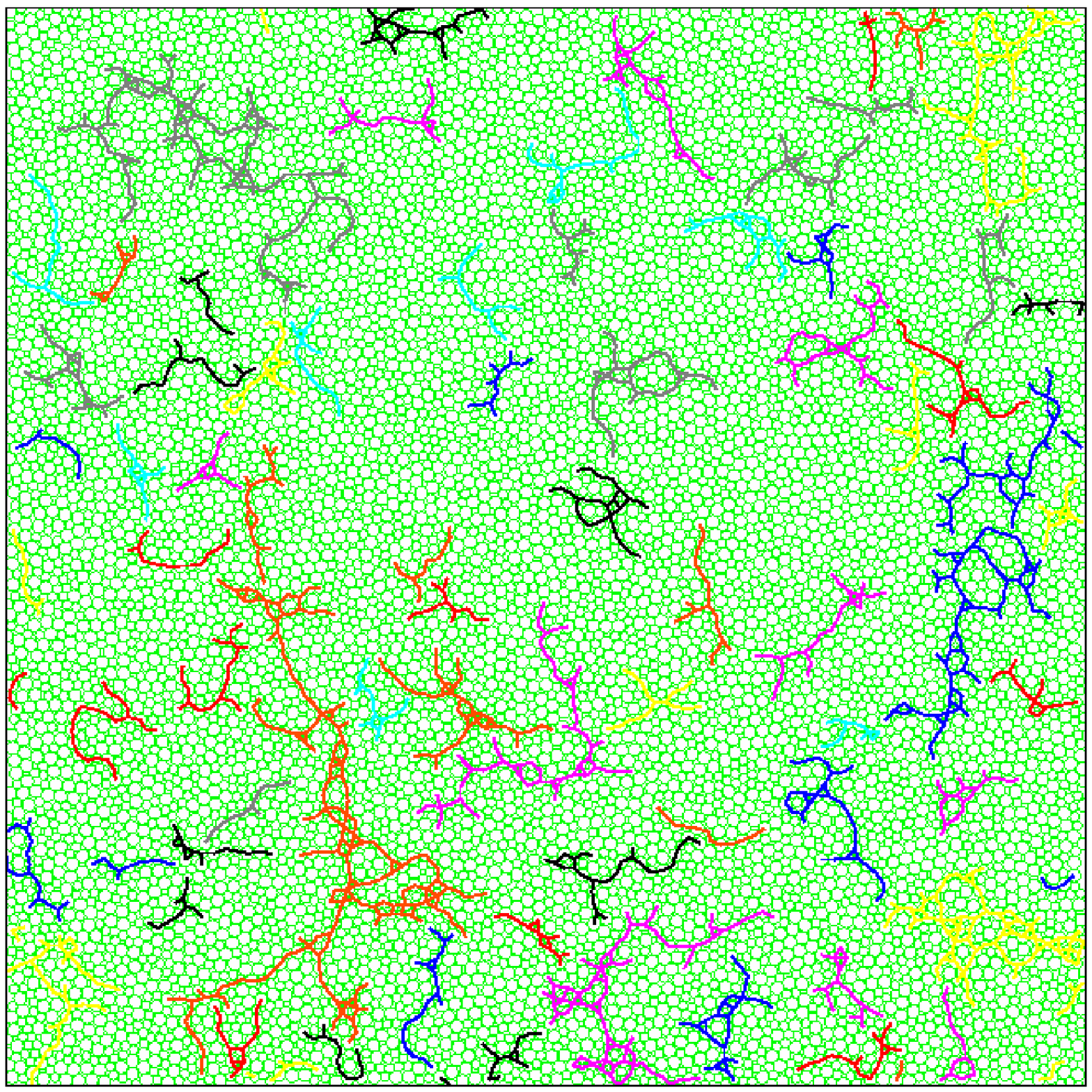}}}
\put(3.5,1.5){\fbox{\includegraphics*[scale=0.13]{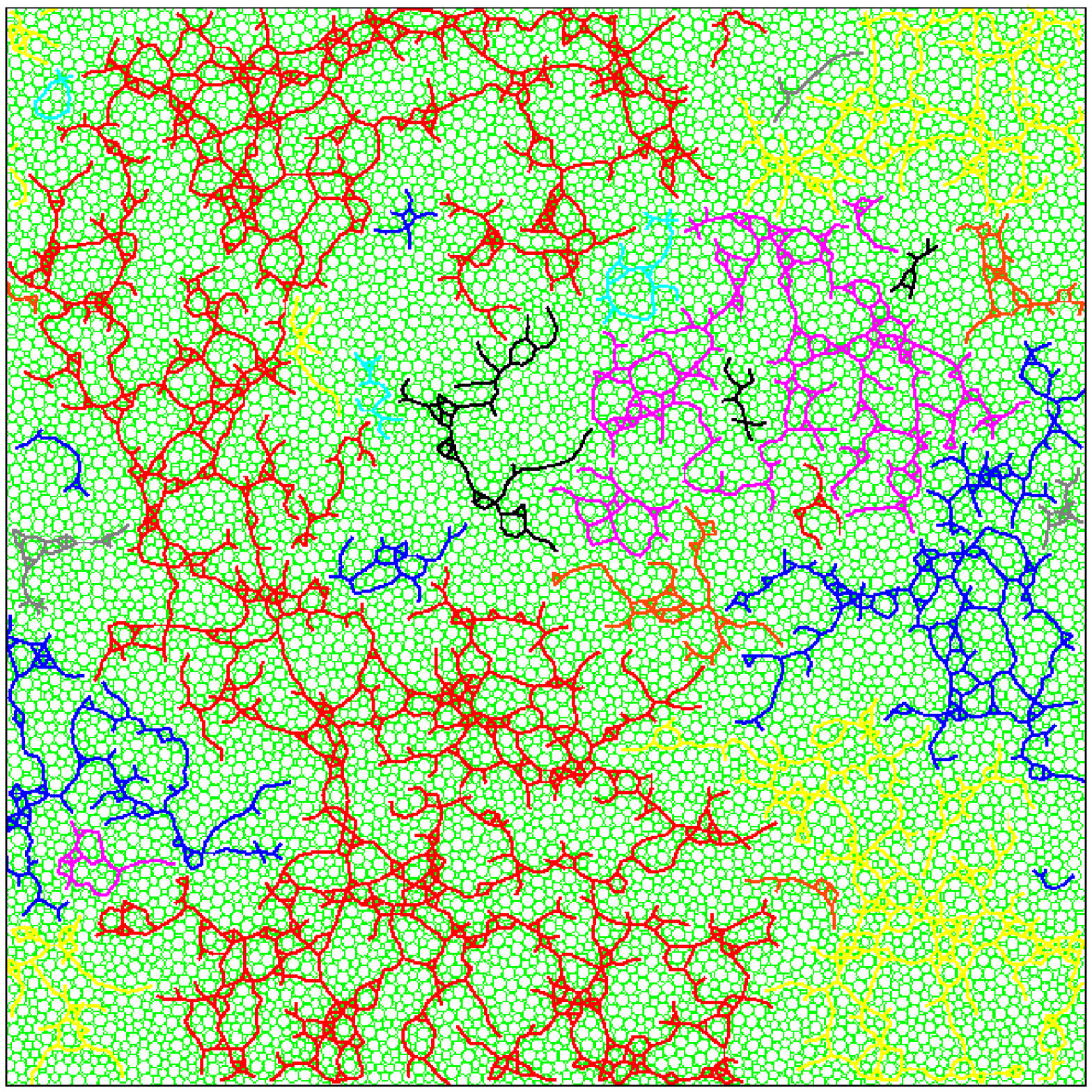}}}
\put(5,0.6){\fbox{\includegraphics*[scale=0.13]{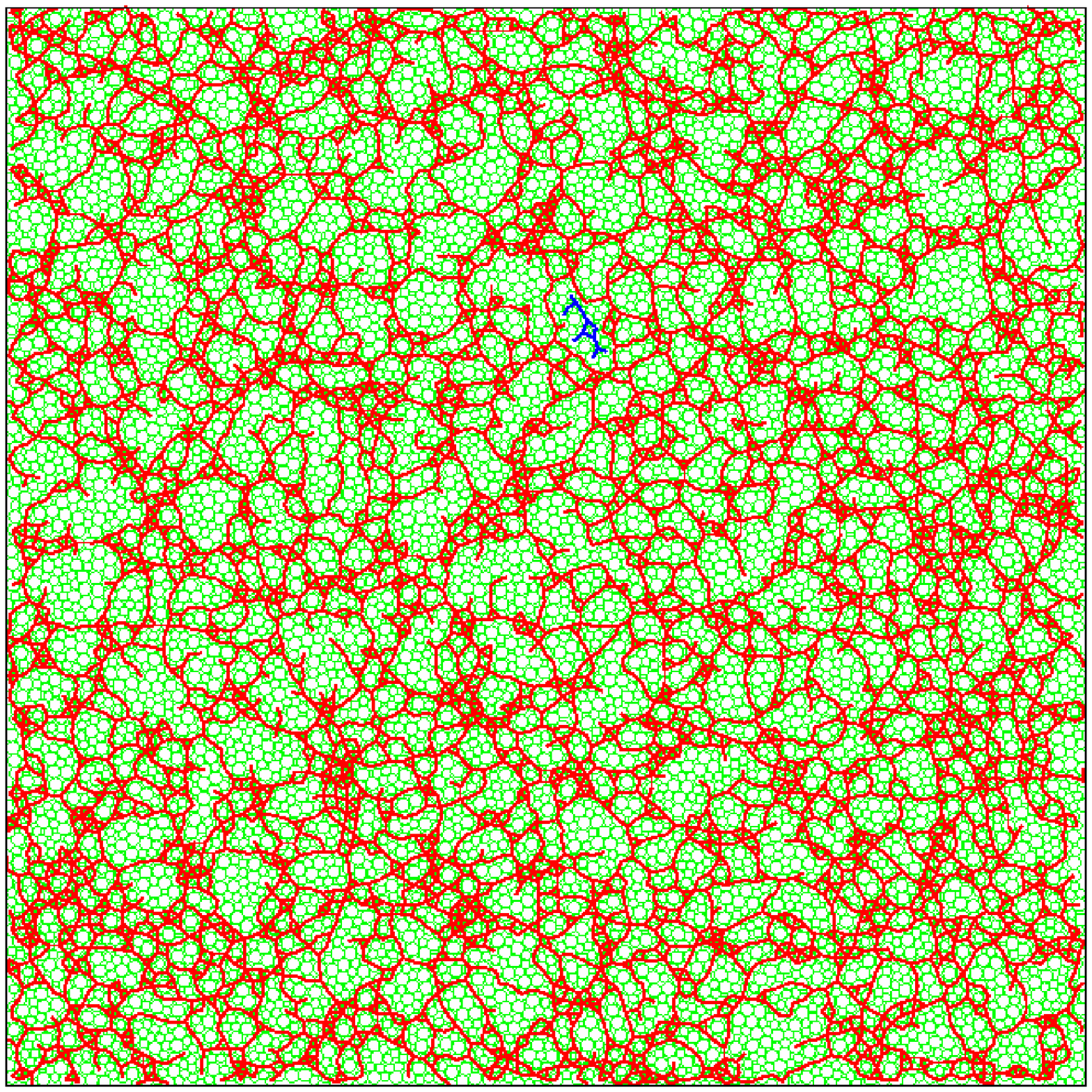}}}
}
\put(1.,1.8){\rotatebox{0}{\rm not Percolating}}
\put(2.5,0.6){\rotatebox{0}{\rm Percolating}}
\end{picture}
\caption{\label{fig:clusters}
A percolative analysis of jammed configurations is introduced by considering particles 
as connected if interacting with a force whose magnitude is greater than a threshold $f_t$. 
This schematic phase diagram illustrates the existence 
of a continuous percolation line that ends at the jamming volume fraction $\Phi_J$, where the 
percolation transition becomes discontinuous. 
The panels illustrate the percolative analysis of a $N = 10^4$ particle system 
across the transition. Lines connect particles belonging to the same cluster.
For clarity, not all of clusters are shown.
} 
\end{figure}

Amorphous particulate systems such as foams and granular materials jam and acquire mechanical 
rigidity when subjected to an external pressure~\cite{CoreyOHern2002PRL,CoreyOHern2003PRE,
HernanMakse2000PRL,TSMajmudar2007PRL,JasnaBrujic2007PRL}.
In the jammed state, a network of interparticle forces determines resistance to 
shear~\cite{DanielHowell1999PRL,MassimoPCiamarra2010PRL}, sound and heat 
transport~\cite{HernanMakse1999PRL,RaulHidalgo2002PRL,ETOwensEPL2011,DanielleSBassett2012PRE}, 
as well as electrical conductivity~\cite{NVandewalleEPL2001}.
The distribution of the magnitude of the interparticle forces has been the subject of 
numerous studies~\cite{FarhangRadjai1996PRL,TSMajmudar2005Nature, KennethDesmond2013SoftMatter,
JZhou2006Science,JasnaBrujic2003Faraday,JasnaBrujic2003PhysicaA,GKatgert2010EPL,
BrianPTighe2010SoftMatter,CoreyOHern2001PRL,JaccoHSnoeijer2004PRL,BrianPTighe2008PRL}, 
and it is now ascertained that this decays exponentially at large forces, while exhibiting 
a pressure dependent power law behavior at small forces. 
Large forces organize along chains~\cite{C-hLiu1995Science,DanielMMueth1998PRE,
FarhangRadjai1999Chaos,JFPeters2005PRE,JZhou2009JSMech,AlexeiVTkachenko2000PRE}, 
which suggests the existence of a large scale structure one might identify through statistical 
physics or network--based tools~\cite{RobertoArevalo2010PRE,LKondic2012EPL,MiroslavKramar2014PRE,
SArdanzaTrevijano2014PRE,DavidMWalker2012PRE,DanielleSBassett2015SoftMatter,
RomualdoPSatorras2012JSMech,Tianqi2012PRE}.
In this line of research, the main open question concerns the 
spatial organization of the force network, 
and the possible existence of long 
range correlations between the forces.
These issues are conveniently investigated studying a force based bond percolation transition in 
which two particles are assumed to belong to a cluster if the magnitude of their interparticle
force is larger than a threshold $f_t$ (see Fig.~\ref{fig:clusters}). 
In the jammed phase, when $f_t = 0$ all contacting particles belong to the same cluster, 
while conversely in the $f_t \to \infty$ limit there are no clusters. 
Thus, a percolation transition occurs when the threshold overcomes a critical value $f_c$.
Ostojic et al.~\cite{SrdjanOstojic2006Nature} numerically investigated this force percolation 
transition in frictionless and frictional systems of disks packings prepared at constant 
pressure, finding a universal critical behavior and exponents not compatible with those of 
the random universality class.
A recent experimental and numerical investigation of the force percolation transition of 
jammed disks packings at constant density~\cite{LKovalcinova2016PRE}
found different critical exponents, also not compatible with the random universality class.
These results point towards the existence of long-range correlations between the forces.
However, direct numerical and experimental investigations of the spatial correlation 
between the forces~\cite{KKarimi2011PRL,KennethDesmond2013SoftMatter} failed to observe 
long correlation lengths.
Accordingly, it is currently unclear whether the correlations between the forces of 
jammed packings are truly long--ranged. 
The answer to this question might depend on the pressure/density
of the system, that controls the percolation threshold $f_c$, as this must vanish
at the jamming transition, as illustrated Fig.~\ref{fig:clusters}, where all forces vanish.
Thus, it is important to understand how the continuous force percolation transition 
in the zero pressure limit relates to the discontinuous jamming transition.

In this Letter, we investigate force correlations in jammed granular packings via the
numerical study of the force percolation transition of Harmonic and Hertzian particles, 
in both two and three spatial dimensions.
First we show that the force percolation transition does actually belong
to the random universality class, regardless of the distance from
the jamming threshold, thus ruling out the presence of long--range
force correlations. Then we clarify how the features of the force percolation 
transition depend on pressure, thus rationalizing how 
the continuous force percolation transition evolves into the discontinuous jamming 
transition in the zero pressure limit. 

{\it Model systems --} We numerically~\cite{StevePlimpton1995JCP} investigate systems of particles interacting via purely repulsive 
potential, $v(r) = \frac{1}{\alpha}\varepsilon\left(\frac{\sigma - r}{\sigma_1}\right)^\alpha$ 
for $r<\sigma$, and $v(r)=0$ for $r \ge \sigma$, where $\varepsilon$ is the characteristic 
energy scale, $\sigma$ is the average diameter of the interacting particles 
and $r$ the distance between their centers.
We study the system for two potentials, $\alpha=2$ (Harmonic) and $\alpha=2.5$ (Hertzian),
focusing on a $50:50$ binary mixture of particles with diameters 
$\sigma_1 = 1$ and $\sigma_2 = \sigma_1/1.4$.
We choose $\sigma_1$ and $\varepsilon$ as our units of length and energy. 
We have considered various systems with number of particles, 
$N = 4\times10^3, 10^4, 2\times10^4, \text{and}\,4\times10^4 $, 
and various values of the pressure, $p = 10^{-2}$, 
$5 \times 10^{-3}$, $2.5 \times 10^{-3}$, $5 \times 10^{-4}$, $5 \times 10^{-5}, 
\text{and} \,\, 5 \times 10^{-6}$. 
To prepare the system at the desired value of pressure,
we first randomly distribute particles in a square/cubic box, with periodic boundary conditions.
The size of the box is then repeatedly changed via a divide and conquer algorithm, where 
we minimize the elastic energy of the system using conjugate gradient algorithm after every 
change in simulation box size.
This iterative procedure continues until the pressure equals the desired value 
with a tolerance of $\left|dp\right|/p < 10^{-6}$.
For each dimensionality $d$, potential and pressure value, our results are averaged 
over $500$ independent jammed configurations.
In this main text, we present results for two dimensional harmonic systems. 
Analogous results for Hertzian potential and the three dimensional system are
presented in supplementary materials~\cite{supplementary}.

{\it Force percolation -- } The force percolation transition is a bond percolation 
transition occurring on a disordered lattice whose nodes correspond to the particles, 
and whose bonds correspond to the interparticle forces.
The percolation is induced by the removal of all bonds associated to interparticle 
forces $f$ lower than the threshold force $f_t$, as the threshold $f_t$ is varied.
At each value of the pressure ($p > 0$) a percolation transition occurs as $f_t$ varies, 
as schematically illustrated in Fig.~\ref{fig:clusters}.
The percolative properties of this transition reflect those of the forces
because the bonds that are retained are not randomly chosen, but correspond to 
interparticle forces greater than $f_t$.

\begin{figure}[h]
\begin{tabular}{c}
\includegraphics[width=\columnwidth]{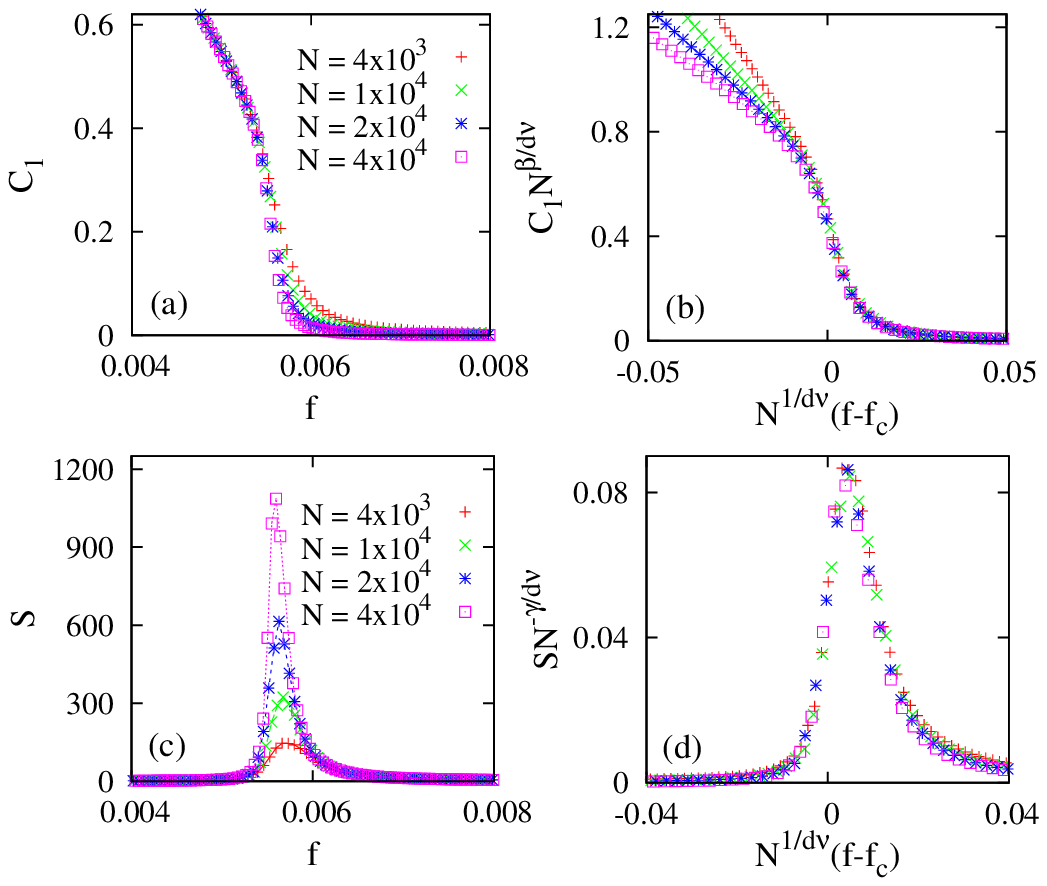}\\
\hspace{-2mm}
\includegraphics[width=0.99\columnwidth]{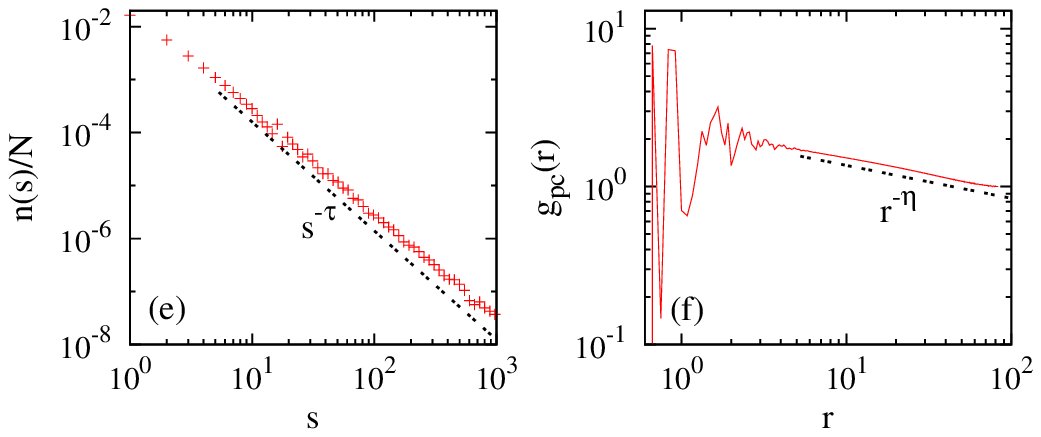}
\end{tabular}
\caption{\label{fig:scaling2dHarmonic}Random percolation universality class of force percolation
transition. Panels (a)-(d) show the finite size scaling of the fraction of particles in the 
largest cluster $C_1$ and mean cluster size $S$ using Eq.~(\ref{eq:size_scaling}). 
The value of $f_c=0.0055$. Panels (e) and (f) show
the normalized cluster size distribution $n(s)$, and the radial distribution function of 
percolating cluster $g_{\text{pc}}(r)$ respectively, at the percolation threshold for 
$N=4\times10^4$ system. Random percolation exponent values for two dimensions, $\nu=4/3$,
$\beta=5/36$, $\gamma=43/18$, $\tau=187/91$, and $\eta=5/24$ are used.
Data is for $p = 5 \times 10^{-3}$.
} 
\end{figure}

We have determined the critical exponents
governing the critical behavior of the strength of the percolating cluster $P_\infty$, of 
the mean cluster size $S$, and of the percolation correlation length $\xi$,
$P_\infty \sim |f-f_c|^{\beta}$,  $S \sim |f-f_c|^{-\gamma}$, $\xi \sim |f-f_c|^{-\nu}$,
performing a standard finite size scaling analysis. Indeed, in finite systems 
the above critical behaviors are replaced by crossovers satisfying
the scaling relations
\begin{eqnarray}
P_\infty(N,f) = N^{-\frac{\beta}{d\nu}} m_1 \left[ N^{\frac{1}{d\nu}} (f-f_c) \right],\nonumber \\
S(N,f) = N^{\frac{\gamma}{d\nu}} m_2 \left[ N^{\frac{1}{d\nu}} (f-f_c)  \right],
\label{eq:size_scaling}
\end{eqnarray}
with $m_1$ and $m_2$ universal scaling functions.
To improve numerical accuracy we have performed
a size scaling analysis of the fraction of particles in largest cluster $C_1$,
that scales as $P_\infty$ but is of easier investigation as
it does not depend on the percolation threshold.
The mean cluster size is defined as $S={\sum s^2n(s)}/{\sum sn(s)}$, where $s$ and $n(s)$ refer
to the size and number of clusters, and the summation excludes the percolating cluster.
The size $s$ of a cluster equals its number of bonds.

Figure~\ref{fig:scaling2dHarmonic} illustrates the finite size scaling 
investigation of the force percolation transition of a two dimensional system of disks
at pressure $p=5\times10^{-3}$. This investigation strongly suggests the percolation 
transition to belong to the random percolation universality class.  
First, panels a-d show our scaling analysis for $C_1$ and $S$. 
Data nicely collapse when rescaled using the random percolation universality class exponents~\cite{DStaufferBook}. 
Second, panels e and f show the normalized cluster size distribution $n(s)$ and the radial 
distribution function of particles belonging to the percolating cluster $g_{\rm pc}(r)$ respectively,
at the percolation threshold for the $N=4\times 10^4$ system.
The distribution $n(s)$ fits very well to the power-law decay $n(s) \sim s^{-\tau}$, 
with $\tau$ Fischer exponent of the random percolation in two dimensions.
Similarly, for large $r$ the pair--connected correlation function is well described 
by a power-law decay $g_{\rm pc}(r) \sim r^{-d+2-\eta}$, with the random 
percolation value of the anomalous dimension $\eta$.
We have found the same results for all values of the pressure 
we have considered.  In addition, analogous findings occur for the Hertzian potential, and in three dimensions, 
as we illustrate in the supplementary material~\cite{supplementary}.
Summarizing, these results clarify that the force percolation
transition is a continuous percolation transition in the random percolation universality class. 
As a consequence, the correlations between the interparticle forces of jammed packings are 
finite ranged.

\begin{figure}[!t]
\includegraphics[width=\columnwidth]{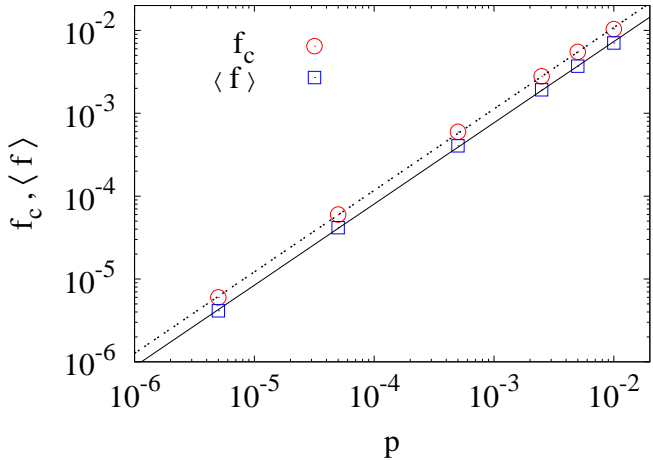}
\caption{\label{fig:cthreshold} Pressure dependence of the force percolation threshold, $f_c$,
and of the average force, $\<f\>$. These two forces are proportional, and scale as 
$f_c = \alpha_c p^q$ and as $\<f\> = \alpha p^q$, 
with $q \approx 0.98(1)$, $\alpha_c=1.00(4)$, and $\alpha=0.67(3)$.}
\end{figure}

It is worth noticing that these results contrast with those
of Refs.~\cite{SrdjanOstojic2006Nature,LKovalcinova2016PRE}, which suggested the force 
percolation transition to be correlated. First, Ref.~\cite{SrdjanOstojic2006Nature} reported
two exponents, $\phi = \frac{\gamma}{d\nu} = 0.89\pm0.01$, and $\nu=1.6\pm0.1$.
Of the two exponents, $\phi$ is compatible with the random percolation expectation, $43/48 = 0.895$,
while $\nu$, which is estimated with lesser accuracy, is not compatible with the random value, $\nu = 4/3$.
We speculate that the difference is due to numerical errors
arising from using small system sizes, as pointed in Ref.~\cite{LKovalcinova2016PRE}. 
Second, Ref.~\cite{LKovalcinova2016PRE} reported $\phi = 0.77-0.85$, and $\nu = 1.04-1.58$
depending on the packing fraction and polydispersity.
Our speculation is that the differences from the random percolation values are due to 
using volume fraction as control parameter.
Indeed, the pressure of jammed packings 
of given volume fraction has strong finite size effects.

\begin{figure}[!t]
\includegraphics[width=\columnwidth]{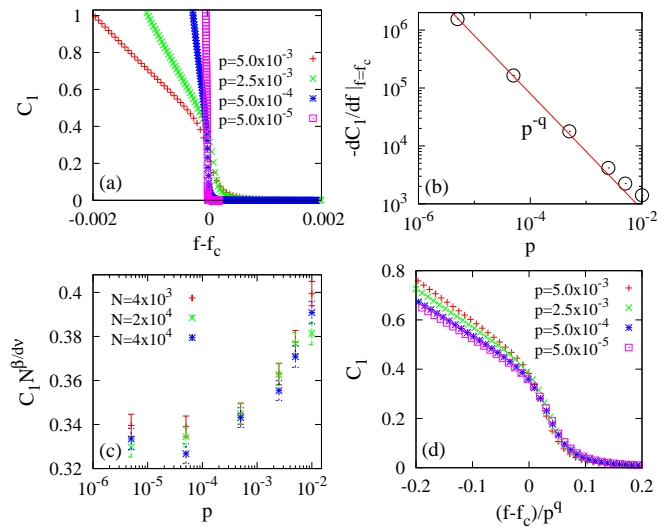}
\caption{\label{fig:inflectionpoint} 
Dependence of the size of the largest cluster, $C_1$, on $f-f_c$, for different values of 
pressure (a), and pressure dependence of its inflection point (b). The derivative diverges 
as $p^{-q}$ in the $p\to0$ limit. 
Panel (c) clarifies that $C_1$ scales as $N^{-\beta/{d\nu}}$, for all values of the pressure,
ruling out the presence of a discontinuous transition in the thermodynamic limit, for $p \to 0$.
Panel (d) shows that the $C_1$ data of (a) collapse when plotted versus $(f-f_c)/p^q$, 
thus clarifying that the size of the critical region scales as $p^{q}$.
In panels a,b and d, $N = 2 \times 10^{4}$.} 
\end{figure}
\begin{figure}[!t]
\includegraphics[width=\columnwidth]{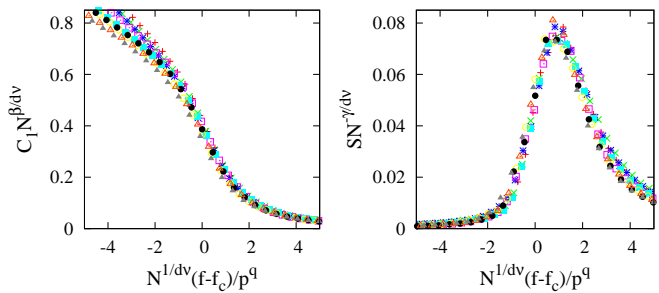}
\caption{\label{fig:combinedscaling} Combined pressure and size scaling of the 
largest cluster size $C_1$ (left panel), and of the mean cluster size $S$ (right panel). 
The presented data are for three system size $N=4\times 10^3,1\times 10^4, \text{and}\, 
2 \times 10^4$, each corresponding to three pressure values $p = 5\times 10^{-4}, 5\times 10^{-5} \text{and}\, 5\times 10^{-6}$. We have fix $q \simeq 0.98$.} 
\end{figure}

We now consider how the force percolation transition, that occurs at fixed pressure
as the threshold force $f_t$ varies, is related to the 
geometrically discontinuous jamming transition
that occurs at $f_t = 0$, as the pressure/density varies.
First, we note that the critical threshold $f_c$ decreases with the pressure
as $f_c \sim p^q$, with $q \approx 0.98$, as illustrated in Fig.~\ref{fig:cthreshold}. 
This is the same scaling we observe for the average force, that is
commonly used~\cite{FarhangRadjai1998PRL} to decompose the force network in 
subnetworks with different mechanical properties.
Next, in Fig.~\ref{fig:inflectionpoint}, we show the pressure dependence of cluster statistics.
We observe in Fig.~\ref{fig:inflectionpoint}a that the size of the largest cluster as function 
of the distance from the percolation threshold $f-f_c$ is pressure dependent, 
and becomes more abrupt when approaching the zero pressure limit.
This tendency is quantified by investigating the derivative of $C_1$ at the 
inflection point, ${dC_1/df |}_{f=f_c}$.
Figure~\ref{fig:inflectionpoint}b shows that this derivative increases in modulus as the pressure 
decreases, diverging as a power-law $\sim p^{-k}$, with $k \simeq q$, in the zero pressure limit. 
This might suggest that the transition becomes discontinuous in the zero pressure limit.
However, we show in Fig.~\ref{fig:inflectionpoint}b that 
$C_1$ at the inflection point has weak dependence on pressure,
and that it scales as $N^{-\frac{\beta}{d\nu}}$. Thus, $C_1$ does not
exhibit a jump of finite size in the $p \to 0$ limit.
Thus, in the zero pressure limit the force percolation transition
remains a continuous transition, and the only effect of the pressure
appears that of controlling the size of the critical region, which is expected to scale as $p^q$. 
We confirm this speculation in Fig.~\ref{fig:inflectionpoint}d,
where we illustrate that the data of panel $a$ collapse when plotted
as a function of $(f-f_c)/p^q$.
Overall, these results suggest a combined 
size and pressure scaling 
for the strength of the percolating cluster, and, similarly, for the mean cluster size,
\begin{eqnarray}
P_\infty(N,p,f) = N^{-\frac{\beta}{d\nu}}m_1 \left[ N^{\frac{1}{d\nu}} p^{-q} (f-f_c) \right],\nonumber \\
S(N,p,f) = N^{\frac{\gamma}{d\nu}}m_2 \left[ N^{\frac{1}{d\nu}} p^{-q} (f-f_c)  \right],
\label{eq:ps_scaling}
\end{eqnarray}
where the exponent $q$ is that controlling the dependence of the 
critical threshold on the pressure (see Fig.~\ref{fig:cthreshold}).
The validity of the proposed scaling relations is confirmed by the good data collapse obtained for various pressure and 
system size, as shown in Fig.~\ref{fig:combinedscaling}.
From Eq.~\ref{eq:ps_scaling}, and the scaling of the percolation threshold on the pressure, 
Fig.~\ref{fig:cthreshold}, we understand that the correlation percolation length,
which measures the typical size of the cluster of forces larger than $f$, scales as 
$\xi = l [p^{-q}(f -f_c)]^{-\nu}$ $= l (\alpha f/\<f\> -\alpha_c)^{-\nu}$, with $l$
pressure independent length scale.
Since in the $p\to 0$ limit the force percolation transition does not
become discontinuous, we understand that the order of the limits $p\to0$ and $f \to 0$ matters.
If the $p \to 0$ limit is carried out first, then the continuous force percolation
transition is observed at $f = 0$. Conversely, if the 
$p \to 0$ limit is carried out first, the one observes the jamming transition 
at $f = 0$.

While these results prove that there are not long-range force correlations,
forces have short range correlations as revealed by
the common observation of force chains. In the percolation setting, short-range correlations
can be revealed comparing the actual percolation threshold $f_c(p)$ to that obtained after 
removing all correlations, $f_c^R(p)$.
The relation between $f_c^R(p)$ and $f_c(p)$ depends on the short-range correlation
length, as well as on anisotropy of the 
correlations~\cite{HMullerKrumbhaar1974PhysLettersA,DStauffer1981JPhysLetters,SASafran1985PRA,MeganEFrary2007PRE}.
We have performed this investigation removing the force correlations by randomly swapping the 
forces associated to the different contacts, treating them as labels,
and have found $f_c^R(p) > f_c(p)$. 
Importantly, we have found $f_c^R(p)/f_c(p)\simeq 0.6$ regardless of the pressure, 
in the $p\to 0$ limit, consistently with our proposed scenario according to which the 
pressure only fixes the size of the critical region.

{\it Conclusions --} 
We have shown that the force percolation transition of jammed
granular packings belongs to the random percolation universality class,
and thus demonstrated the absence of long-ranged force correlations,
contrary to earlier speculations~\cite{SrdjanOstojic2006Nature}. 
This result occurs regardless of the distance from the jamming transition,
as proved by a combined size and pressure scaling.
The main peculiarity of this force percolation transition is the dependence
of the width of the critical region on the pressure, and thus on percolation threshold,
as in the jamming zero pressure limit the critical region disappear.
While we do expect this scenario to hold regardless of the protocol used to prepare
the jammed packings, and thus regardless of the jamming volume 
fraction~\cite{MPicaCiamarra2010SoftMatter,PinakiChaudhuri2010PRL}, 
this is certainly an interesting avenue of research. 
Similarly, it would be of interest to investigate force correlations
on the unjammed side of the transition, where interparticle forces can be defined
in hard sphere systems from the collisional momentum exchange~\cite{AleksandarDonev2005PRE}. 
We finally remark that, while we have shown that two--body force correlations are random in nature,
regardless of the jamming transition, recent results~\cite{MMailman2012JSM} have shown the 
existence of a point--to--set force correlation length that diverges at jamming.

\begin{acknowledgments}
MPC acknowledge support from the Singapore Ministry of Education through the Academic Research Fund (Tier 1) 
under Projects No. RG104/15 and RG179/15. We thank E. Weeks for discussions.
\end{acknowledgments}

\end{document}